\documentclass[%
 reprint,
 superscriptaddress,preprintnumbers,
 nofootinbib,
 amssymb,
 aps,
]{revtex4-2}

\usepackage[utf8]{inputenc}
\usepackage{hyperref}
\usepackage{amsmath,amssymb,mathtools}
\usepackage{dsfont}
\usepackage{graphicx}   
\usepackage[table]{xcolor}
\usepackage{colortbl}
\usepackage{url}
\usepackage[normalem]{ulem}
\usepackage{slashed}
\usepackage[capitalise, english]{cleveref}
\usepackage[noindentafter]{titlesec} 
\allowdisplaybreaks

\usepackage{siunitx}
\usepackage{xspace}
\usepackage{enumitem}

    \makeatletter
    \def\CT@@do@color{%
      \global\let\CT@do@color\relax
            \@tempdima\wd\z@
            \advance\@tempdima\@tempdimb
            \advance\@tempdima\@tempdimc
    \advance\@tempdimb\tabcolsep
    \advance\@tempdimc\tabcolsep
    \advance\@tempdima2\tabcolsep
            \kern-\@tempdimb
            \leaders\vrule
                    \hskip\@tempdima\@plus  1fill
            \kern-\@tempdimc
            \hskip-\wd\z@ \@plus -1fill }
    \makeatother

\usepackage{accents}
\DeclareMathSymbol{\widetildesym}{\mathord}{largesymbols}{"65}

\def\thesubsection{\arabic{section}.\arabic{subsection}}

\def\thesection{\arabic{section}}
\titleformat*{\subsubsection}{\normalfont \small \bfseries \boldmath}
\renewcommand{\paragraph}[1]{\vspace{.3em} \indent {\bfseries \boldmath #1 ---}\xspace }
\makeatletter
    \renewcommand{\p@subsection}{}
    \renewcommand{\p@subsubsection}{}
\makeatother

\definecolor{red}{rgb}{0.6,.0706,.1373}
\definecolor{blue}{rgb}{0,0.396,0.741}
\newcommand\myshade{80}
\colorlet{mylinkcolor}{violet}
\colorlet{mycitecolor}{violet}
\colorlet{myurlcolor}{violet}

\hypersetup{
  linkcolor  = mylinkcolor!\myshade!black,
  citecolor  = mycitecolor!\myshade!black,
  urlcolor   = myurlcolor!\myshade!black,
  colorlinks = true
}
\newcommand{\phiH}{\phi_{21}}
\newcommand{\phiL}{\phi_{32}}
\newcommand{\LamH}{\Lambda_{21}}
\newcommand{\LamL}{\Lambda_{32}}

\newcommand{\U}{\mathrm{U}}
\newcommand{\SU}{\mathrm{SU}}


\begin{document}





\title{
 \boldmath Neutrino Anarchy from Flavor Deconstruction
}

\author{Admir Greljo}
\email{admir.greljo@unibas.ch}
\affiliation{Department of Physics, University of Basel, Klingelbergstrasse 82,  CH-4056 Basel, 
Switzerland}

\author{Gino Isidori}
\email{isidori@physik.uzh.ch}
\affiliation{Physik-Institut, Universität Zürich, CH-8057 Zürich, Switzerland}



\begin{abstract}
Flavor deconstruction refers to non-universal gauge extensions where the original gauge symmetry is deconstructed into separate copies, one for each family. A hierarchical chain of symmetry breaking provides an attractive low-scale solution to the flavor puzzle, consistent with flavor-changing neutral currents and finite naturalness. Although successful in explaining the origin of flavor hierarchies in the quark and charged lepton sectors, existing models have struggled with the large and seemingly anarchic mixing observed in neutrino oscillations. This letter identifies conditions under which neutrino anarchy may arise from flavor deconstruction in generic models with right-handed neutrinos. When deconstruction is applied to carefully chosen subgroups of the extended gauge symmetry, hierarchies in the seesaw formula cancel out.
\end{abstract}

\maketitle

\section{Introduction} 
\label{sec:intro}

Flavor deconstruction refers to a framework wherein a gauge group $ G $ is extended in the ultraviolet to $G^3$, incorporating one $G$ factor for each family. Such non-universal gauge extensions of the Standard Model (SM) have gained significant attention recently~\cite{Bordone:2017bld, Greljo:2018tuh, Allanach:2018lvl, Greljo:2019xan, Fuentes-Martin:2020bnh, Allwicher:2020esa, Fuentes-Martin:2020pww,  Davighi:2022fer, Davighi:2022bqf, Davighi:2023iks, Davighi:2023evx, FernandezNavarro:2023rhv,  FernandezNavarro:2023hrf, Davighi:2023xqn, Barbieri:2023qpf,  Capdevila:2024gki, Fuentes-Martin:2024fpx, FernandezNavarro:2024hnv}. The chain of spontaneous symmetry breaking initially proceeds through $G_1 \times G_2 \to G_{12}$, facilitated by a scalar link field $\phiH$ acquiring a non-zero vacuum expectation value (VEV) $\langle \phiH \rangle$. This is followed by $G_{12} \times G_{3} \to G$, mediated by another scalar link field $\phiL$, which connects $G_{2} \times G_{3}$. Fermions $ f_i $ are individually charged under the respective $ G_i $ factor associated with each flavor $ i = 1, 2, 3 $.\footnote{Since the anomaly cancelation takes place within a family, such gauge extensions are automatically anomaly-free.} 

Spontaneous symmetry breaking to a diagonal subgroup for simple groups is robustly predicted for a nontrivial scalar representation under both factors. However, it is economical to choose link fields such that $\phiH f_1$ and $f_2$ belong to the same gauge representation, likewise for $\phiL f_2$ and $f_3$. In addition, the Higgs field $H$ is only charged under $G_3$. This sets the basis for explaining the flavor hierarchies: the inter-family mass rations are  controlled by the suppression factors $\epsilon_1=\langle \phiH \rangle / \LamH$
and $\epsilon_2=\langle \phiL \rangle / \LamL$, of $O(10^{-2})$, where $\Lambda_{ij}$ generically denote the scales of the effective operators yielding at low energies the SM Yukawa couplings. In addition, separating the light family dynamics from the third family by $\langle \phiH \rangle \gg \langle \phiL \rangle \gg \langle H \rangle$ allows for a low-scale flavor model consistent with the flavor-changing neutral currents and finite naturalness~\cite{Davighi:2023iks}.\footnote{Flavor deconstruction could originate from a warped extra dimension, see e.g.~\cite{Fuentes-Martin:2022xnb}.}

While hierarchical singular values are observed in the quark and charged-lepton mass matrices, the neutrino mass matrix appears to be anarchic. In contrast to the small and hierarchical mixing between quark generations, neutrino oscillation experiments reveal large $\mathcal{O}(1)$ mixing angles, see e.g.~\cite{Esteban:2020cvm, Capozzi:2017ipn}. Even the ratio of mass differences $|\Delta m^\nu_{\rm atm}|/|\Delta m^\nu_{\rm sol}| \approx 5$ is not particularly large compared to the mass hierarchies in the charged fermion sector between consecutive generations, which are on average $\mathcal{O}(100)$. The question addressed in this letter is whether and under which conditions flavor deconstruction can lead to flavor anarchy in the neutrino sector together with the observed hierarchical Yukawa couplings
for quarks and charged leptons.


This question cannot be addressed without making hypotheses about the origin of neutrino masses. We analyze the problem under the general assumption that the set of chiral fermions is extended by at least three right-handed neutrinos. Under this assumption,  we first analyze in detail the representative case of the type-I seesaw mechanism (Section~\ref{sec:typeI}). We then show how the conclusions derived in that case can easily be generalized in the presence of an inverse seesaw mechanism (Section~\ref{sec:typeII}).

\section{Deconstructing the type-I seesaw} 
\label{sec:typeI}

 In this section, we will limit ourselves to the type-I seesaw mechanism~\cite{Minkowski:1977sc, Yanagida:1980xy, Gell-Mann:1979vob, Mohapatra:1979ia, CentellesChulia:2024uzv}, where 
 neutrino masses are described by the following Lagrangian
\begin{equation}
   - \mathcal{L} \supset Y^{ij}_\nu \bar \ell_i \tilde H \nu_j + \frac{1}{2} M_{\rm M}^{i j} \nu_i \nu_j + {\rm h.c.}
\end{equation}
Here, $\ell_i$ and $\nu_i$ are the left-handed lepton doublets and right-handed neutrinos, while $\tilde H = i \sigma_2 H^*$. After integrating out heavy $\nu_i$, the mass matrix for active neutrinos becomes
\begin{equation}\label{eq:typeI}
    m_\nu \approx - \langle H \rangle^2~ Y_\nu \,M_{\rm M}^{-1} (Y_\nu)^T ~ .
\end{equation}
To understand how flavor deconstruction acts on $m_\nu$, we need to specify which gauge group is deconstructed. It is easy to realize that deconstructing any of the three simple subgroups of the  SM gauge symmetry prevents anarchic neutrino mixing and hierarchical quark and charged-lepton masses simultaneously. More precisely, producing the desired $Y_{u, d, e}$ singular value hierarchies requires deconstructing either $ \SU(2)_{\rm L}$ or $ \U(1)_{\rm Y}$ or both.\footnote{The deconstruction of QCD, which is a vector-like group, does not predict mass hierarchies.} This inevitably imprints hierarchies in $ Y_\nu$ while keeping $M_{\rm M}$ intact since $\nu_i$ are SM gauge singlets. In particular,
\begin{equation}\label{eq:eq3}
Y_{\nu} \sim \begin{bmatrix}
\epsilon^2 & \epsilon^2 & \epsilon^2 \\
\epsilon & \epsilon & \epsilon \\
1 & 1 & 1
\end{bmatrix}~,
\end{equation}
for both $ \SU(2)_{\rm L}$ and $ \U(1)_{\rm Y}$ deconstruction, where we assumed $\epsilon_1 = \epsilon_2 = \epsilon$ for simplicity. In this case, diagonalising the complex  matrix $m_\nu$ in Eq.~\eqref{eq:typeI} via $U_\nu^T m_\nu U_\nu =  \hat m_\nu $, leads to $U_\nu \approx {\bf 1}$. The left-handed charged lepton mixing has a similar form, implying a neutrino mixing matrix~\cite{Maki:1962mu} close to the identity: $V_{\rm PMNS} = U_\nu^\dagger U_{e_L}\approx {\bf 1}$.\footnote{This is not specific to the type-I seesaw mechanism and can be formulated at the level of the Weinberg operator $ \mathcal{L} \supset \frac{c_{ij}}{\Lambda} \ell_i \ell_j H H $. Since $\ell_i$ are charged under both $ \SU(2)_{\rm L}$ and $ \U(1)_{\rm Y}$, $c_{ij}$ is necessarily hierarchical under their deconstruction. Obviously, the same holds if neutrinos are Dirac, see Eq.~\eqref{eq:eq3}, though this case is less motivated as it does not address the overall smallness of neutrino masses. }

A necessary ingredient to achieve our goal is an extended gauge group where the \(\nu_i\) are not singlets, allowing \(M_{\rm M}\) to acquire a non-trivial flavor pattern that can compensate for the (unavoidably) hierarchical structure of \(Y_\nu\). Without loss of generality, given the fermion content under consideration, we enlarge the gauge group to
\begin{equation}
\SU(3)_{\rm C} \times \SU(2)_{\rm L} \times \U(1)_{\rm R} \times \U(1)_{\rm B-L}\,,
\label{group}
\end{equation}
where hypercharge is a subgroup of the two $\U(1)$ factors:\footnote{Both $\U(1)$ groups can be embedded in motivated non-Abelian groups; however, this makes no difference for the present discussion.}
$\rm Y = R + (B - L)/2$. In particular, the right-handed neutrinos carry ${\rm R} = +1/2$ and ${\rm L} = 1$ such that $\rm Y = 0$. 
The full deconstruction of either $\U(1)_{\rm R}$ or $\U(1)_{\rm B-L}$, with the other remaining universal, leads to the following texture for the Majorana mass matrix:
\begin{equation}\label{eq:majorana}
M_{\rm M} \sim \begin{bmatrix}
\epsilon^4 & \epsilon^3 & \epsilon^2 \\
\epsilon^3 & \epsilon^2 & \epsilon \\
\epsilon^2 & \epsilon & 1
\end{bmatrix}~.
\end{equation}
In order to get an anarchic $m_\nu$, the following form of the Dirac mass matrix is required (Appendix~\ref{app:typeI}):
\begin{equation}\label{eq:condition}
Y_{\nu} \sim \begin{bmatrix}
\epsilon^2 & \lesssim \epsilon &  \lesssim 1 \\
\lesssim \epsilon^2 & \epsilon & \lesssim 1 \\
\lesssim \epsilon^2 & \lesssim \epsilon & 1
\end{bmatrix}~.
\end{equation}
No entry in a given column should be parametrically larger than the diagonal one. Additionally, the diagonal entries must strictly adhere to the $(\epsilon^2, \epsilon, 1)$ power counting.

The key to a successful description of neutrino, charged-lepton, and quark mass matrices is  deconstructing some of the simple groups in Eq.~\eqref{group} in order to achieve 
both Eqs.~\eqref{eq:majorana} and \eqref{eq:condition}, together with hierarchical (and properly aligned) $Y_{u,d,e}$.

The Majorana mass matrix assumes the inverse hierarchical form under the following three hypotheses:
\begin{enumerate}[label=\Alph*)]
    \item $\U(1)^3_{\rm R} \times \U(1)_{\rm B-L}$,
    \item $\U(1)_{\rm R} \times \U(1)^3_{\rm B-L}$, 
    \item  $\U(1)^3_{\rm R} \times \U(1)^3_{\rm B-L}$.
\end{enumerate}
In the following, we proceed to analyze these three options in detail. As we will show, all of them can lead to a phenomenologically viable pattern for all mass matrices if accompanied by suitable deconstructions of either $\SU(3)_C$
(option A) or $\SU(2)_{\rm L}$ (options B and C).

\subsection{Model A: $\SU(3)^2_{\rm C} \times \SU(2)_{\rm L} \times \U(1)^3_{\rm R} \times \U(1)_{\rm B-L}$}
\label{sec:modelI}

\begin{figure*}
  \centering
  \includegraphics[trim={0.2cm 0cm 0cm 0cm}, clip, width=0.8\textwidth]{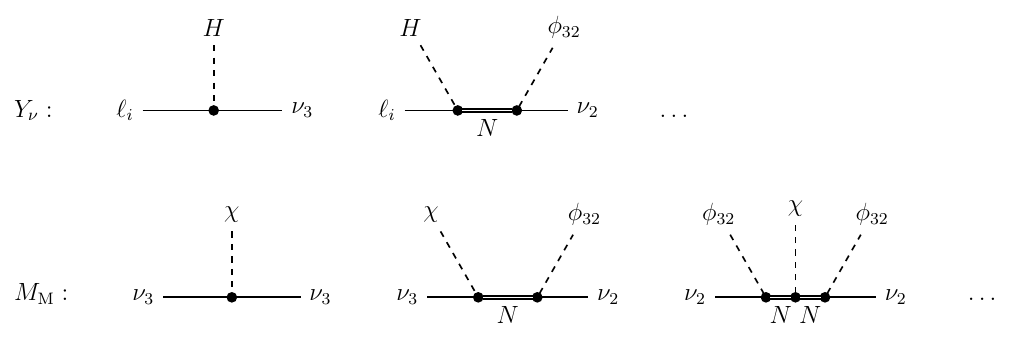}
  \caption{A representative UV completion of selected operators in Eqs.~\eqref{eq:opD} and \eqref{eq:opM}. A heavy vectorlike fermion $N = N_L \oplus N_R$ in the gauge representation of $\nu_3$, and mass $m_N \gg \langle \chi \rangle, \langle \phi_i \rangle$, when integrated out, ensures the same suppression factor $\epsilon_2 = \langle \phi_2 \rangle / m_{N}$ in $Y_{\nu}$ and $M_{\rm M}$ matrices. Similarly, adding another vectorlike fermion in the gauge representation of $\nu_2$ allows the completion of the first-family structure. }
  \label{fig:diagram}
\end{figure*}

The scalar field content of this option is illustrated in Table~\ref{tab:scalars1}. Note its minimality, leaving no uneaten Goldstone bosons.
\begin{table}[t]
\centering
\renewcommand{\arraystretch}{1.5}
\begin{tabular}{|c|c|c|c|c|}\hline
\rowcolor{black!15}[] Scalars & \( \U(1)^{(1)}_{\rm R} \) & \( \U(1)^{(2)}_{\rm R} \) & \( \U(1)^{(3)}_{\rm R} \) & \( \U(1)_{\rm B-L} \) \\ 
\hline
\( \phiH \) & \( 1/2 \) & \( -1/2 \) & \( 0 \) & \( 0 \) \\
\( \phiL \) & \( 0 \) & \( 1/2 \) & \( -1/2 \) & \( 0 \) \\ \hline
\( \chi \) & \( 0 \) & \( 0 \) & \( -1 \) & \( +2 \) \\
\( H \) & \( 0 \) & \( 0 \) & \( 1/2 \) & \( 0 \) \\ \hline
\end{tabular}
\caption{Charges of the symmetry-breaking scalars under \( \U(1)^3_{\rm R} \times \U(1)_{\rm B-L} \).}
\label{tab:scalars1}
\end{table}
The effective theory operators generating the leptonic Yukawa interactions are
\begin{equation}\label{eq:opD}
\begin{split}
-\mathcal{L}_Y \supset &\quad \overline{\ell}_{i} H e_{3} + \frac{1}{\LamL}\overline{\ell}_{i} H \phiL e_{2} + \frac{1}{\LamL \LamH}\overline{\ell}_{i} H \phiL \phiH e_{1} \\
&+ \overline{\ell}_{i} \tilde{H} \nu_{3} + \frac{1}{\LamL}\overline{\ell}_{i} \tilde{H} \phiL^{*} \nu_{2} + \frac{1}{\LamL \LamH}\overline{\ell}_{i} \tilde{H} \phiL^{*} \phiH^{*} \nu_{1}~,
\end{split}
\end{equation}
where we have omitted to show $\mathcal{O}(1)$ parameters in front of the operators. Similarly, the Majorana mass matrix gets generated from
\begin{equation}\label{eq:opM}
\begin{split}
-\mathcal{L}_M = &\quad \chi \nu_3 \nu_3 + \frac{1}{\LamL} \chi \phiL^{*} \nu_3 \nu_2 + \frac{1}{\LamL \LamH} \chi \phiL^{*} \phiH^{*} \nu_3 \nu_1 \\
&+ \frac{1}{\LamL^2} \chi \phiL^{*} \phiL^{*} \nu_2 \nu_2 + \frac{1}{\LamL^2 \LamH} \chi \phiL^{*} \phiL^{*} \phiH^{*} \nu_2 \nu_1 \\ 
&+ \frac{1}{\LamL^2 \LamH^2} \chi \phiL^{*} \phiL^{*} \phiH^{*} \phiH^{*} \nu_1 \nu_1 ~.
\end{split}
\end{equation}
The common effective scales $\LamH$ and $\LamL$ lead to
the universal suppression factors $\epsilon_i = \langle \phi_{ij} \rangle / \Lambda_{ij}$.\footnote{A straightforward UV completion which ensures the effective neutrino operators in Eqs.~\eqref{eq:opD} and \eqref{eq:opM} are suppressed by the same scales $\LamH$ and $\LamL$ involves a tree-level mediation of heavy vector-like fermions that transform in the same gauge representations as $\nu_2$ and $\nu_3$, respectively. See Fig.~\ref{fig:diagram}.} For simplicity, we assume universal $\epsilon_1 = \epsilon_2 = \epsilon$ as before. 

The deconstruction of $\U(1)_{\rm R}$ leads to Yukawa couplings of the form
\begin{equation} \label{eq:U1Rdec}
Y_{u,d,e,\nu}^{\U(1)^3_{\rm R}} \sim \begin{bmatrix}
\epsilon^2 & \epsilon & 1 \\
\epsilon^2 & \epsilon & 1 \\
\epsilon^2 & \epsilon & 1
\end{bmatrix}~,
\end{equation}
while the Majorana mass matrix takes the form in Eq.~\eqref{eq:majorana}. The singular values of $Y_{e}$ are $(\epsilon^2, \epsilon, 1)$ producing the observed mass hierarchies. Furthermore, given the condition in Eq.~\eqref{eq:condition} is satisfied, the resulting neutrino mass matrix $m_\nu$ is anarchic.

However, the left-handed mixing matrices obtained from the singular value decomposition have $\mathcal{O}(1)$ angles. While this works well for $V_{\rm PMNS}$, it is not a viable structure for $V_{\rm CKM}$~\cite{Cabibbo:1963yz, Kobayashi:1973fv} obtained from the left-handed mixing in the quark sector. This problem can be addressed through the flavor deconstruction of QCD without compromising the leptonic sector. Note that since QCD is a vector-like group, its deconstruction does not affect quark masses that remain similar to the charged-lepton ones. 

Consider, for concreteness, a partial QCD deconstruction $\SU(3)^{(12)}_{\rm C} \times \SU(3)^{(3)}_{\rm C}$. A scalar field $\Omega \sim ({\bf 3}, {\bf \bar 3})$ breaks the symmetry down to the QCD. The effective Lagrangian for the quark sector becomes
\begin{equation}
\begin{split}
-\mathcal{L}_Y \supset &\quad  \overline{q}_{3} H d_{3} + \frac{1}{\Lambda'} \overline{q}_{p} H \Omega d_{3}
\\ &+ \frac{1}{\LamL \Lambda'}\overline{q}_{3} H \phiL \Omega^\dagger d_{2} + \frac{1}{\LamL \LamH \Lambda'}\overline{q}_{3} H \phiL \phiH \Omega^\dagger d_{1} \\ 
&+ \frac{1}{\LamL}\overline{q}_{p} H \phiL d_{2} + \frac{1}{\LamL \LamH}\overline{q}_{p} H \phiL \phiH d_{1} \\
&+\overline{q}_{3} \tilde H u_{3} + \frac{1}{\Lambda'} \overline{q}_{p} \tilde H \Omega u_{3}
\\ &+ \frac{1}{\LamL \Lambda'}\overline{q}_{3} \tilde H \phiL^{*} \Omega^\dagger u_{2} + \frac{1}{\LamL \LamH \Lambda'}\overline{q}_{3} \tilde H \phiL^{*} \phiH^{*} \Omega^\dagger u_{1} \\ 
&+ \frac{1}{\LamL}\overline{q}_{p} \tilde H \phiL^{*} u_{2} + \frac{1}{\LamL \LamH}\overline{q}_{p} \tilde H \phiL^{*} \phiH^{*} u_{1}~.
\end{split}
\end{equation}
Here, $p=1,2$ and $\Lambda'$ is the effective scale of operators containing insertions of $\Omega$. The corresponding suppression factor $\epsilon' = \langle \Omega \rangle / \Lambda$ shows up only in the mixing. The quark Yukawa matrices, therefore, take the form
\begin{equation}
Y^{\rm I}_{u,d} \sim \begin{bmatrix}
\epsilon^2 & \epsilon  & \epsilon' \\
\epsilon^2 & \epsilon & \epsilon' \\
\epsilon' \epsilon^2  & \epsilon' \epsilon & 1
\end{bmatrix}~.
\end{equation}
The singular values of $Y_{u,d}$ are $(\epsilon^2, \epsilon, 1)$ in agreement with the charged leptons. The predicted Cabbibo angle is large while $V_{cb}$ and $V_{ub}$ are of order $\epsilon'$. A full deconstruction of QCD would predict $V_{us} \sim \epsilon'_1$, $V_{cb} \sim \epsilon'_2$ and $V_{ub} \sim V_{us} V_{cb}$.

A relevant question to address is the identification of the smallest scale at which deconstruction occurs. The EFT description in Eq.~\eqref{eq:opM} is valid only when $\Lambda_{1,2} \gtrsim \langle \chi \rangle$. The absolute neutrino mass scale naturally points to $\langle \chi \rangle \approx \langle H \rangle^2 / (0.1\,\rm{eV}) \sim 10^{14}$\,GeV. On the other hand, the charged fermion hierarchies are best fitted for $\epsilon \sim 4\times 10^{-3}$ (see the Supplemental Material of~\cite{Antusch:2023shi}). Assuming the same $\epsilon$ for neutrinos and charged leptons, this points to a high scale for the $\U(1)_{\rm R}$ deconstruction, which makes it challenging for experiments. Moreover, this scenario would require a huge tuning of the Higgs mass term in the absence of a protection mechanism such as low-energy supersymmetry or compositeness.


One way to lower the scale would be to introduce a universal suppression factor in the $Y_\nu$ matrix. A possible option is a flavor-universal $Z_2$ symmetry under which only the $\nu_i$ are odd. Denoting $\mu_\nu$ the corresponding symmetry-breaking spurion, $\langle \chi \rangle \sim (\mu_\nu/10^{-4})^2 \times 1000$~TeV. A more motivated option is to implement flavor deconstruction in a symmetry-protected seesaw mechanism, such as the inverse seesaw. We discuss this option in Section~\ref{sec:typeII}.

\subsection{Model B: $\SU(3)_{\rm C} \times \SU(2)^3_{\rm L} \times \U(1)_{\rm R} \times \U(1)^3_{\rm B-L}$}
\label{sec:modelII}

A minimal option to 
deconstruct  $\U(1)_{\rm B-L}$ and generate the effective SM Yukawa couplings is via the scalar field content in Table~\ref{tab:scalars2}.
This way, all Goldstone bosons are gauged away. However, one can consider less minimal options also introducing the link fields $\phi^\ell_{ij} \sim (\phi_{ij}^{q*})^3$. This is why, to keep the discussion more general, we define $\epsilon^{q,\ell}_i = \langle \phi^{q,\ell}_{ij} \rangle / \Lambda^{q,\ell}_{ij}$.
\begin{table}[t]
\centering
\renewcommand{\arraystretch}{1.5}
\begin{tabular}{|c|c|c|c|c|}\hline
\rowcolor{black!15}[]
Scalars & \( \U(1)^{(1)}_{\rm B-L} \) & \( \U(1)^{(2)}_{\rm B-L} \) & \( \U(1)^{(3)}_{\rm B-L} \) & \( \U(1)_{\rm R} \) \\ 
\hline
\( \phiH^q \) & \( -1/3 \) & \( 1/3 \) & \( 0 \) & \( 0 \) \\
\( \phiL^q \) & \( 0 \) & \( -1/3 \) & \( +1/3 \) & \( 0 \) \\ \hline
\( \chi \) & \( 0 \) & \( 0 \) & \( +2 \) & \( -1 \) \\
\( H \) & \( 0 \) & \( 0 \) & \( 0 \) & \( 1/2 \) \\ \hline
\end{tabular}
\caption{Charges of the minimal set of symmetry-breaking scalars under $ \U(1)_{\rm R} \times \U(1)^3_{\rm B-L} $.}
\label{tab:scalars2}
\end{table}
With these assumptions, $\U(1)^3_{\rm B-L}$ implies
\begin{equation}
Y^{\U(1)^3_{\rm B-L}}_{u,d} \sim \begin{bmatrix}
1 & \epsilon_q &  \epsilon_q^2 \\
\epsilon_q & 1 & \epsilon_q \\
\epsilon_q^2 & \epsilon_q & 1
\end{bmatrix},\quad
Y^{\U(1)^3_{\rm B-L}}_{e,\nu} \sim \begin{bmatrix}
1 & \epsilon_\ell &  \epsilon_\ell^2 \\
\epsilon_\ell & 1 & \epsilon_\ell \\
\epsilon_\ell^2 & \epsilon_\ell & 1
\end{bmatrix},
\end{equation}
where in the minimal model $\epsilon_\ell = \epsilon^3_q$.
The Majorana mass matrix takes the form in Eq.~\eqref{eq:majorana} with the replacement $\epsilon \to \epsilon_\ell$. This texture does not generate neutrino anarchy nor charged fermion mass hierarchy, calling for an additional deconstruction of the 
$\SU(2)_{\rm L}$ group. To this purpose, we introduce another pair of scalar link fields, bi-doublets under $\SU(2)^{(1)}_L \times \SU(2)^{(2)}_L $ and $\SU(2)^{(2)}_L \times \SU(2)^{(3)}_L $, respectively. 
The Higgs is charged under 
$ \SU(2)^{(3)}_L $. Denoting $\epsilon_L$ the 
suppression factor associated to the 
$\SU(2)_{\rm L}$ deconstruction, one has
\begin{equation}
Y^{\SU(2)^3_{\rm L}}_{u,d,e,\nu} \sim \begin{bmatrix}
\epsilon_{\rm L}^2 & \epsilon_{\rm L}^2 &  \epsilon_{\rm L}^2 \\
\epsilon_{\rm L} & \epsilon_{\rm L} & \epsilon_{\rm L} \\
1 & 1 & 1
\end{bmatrix}~.
\end{equation}
Hence the combined effect of $\SU(2)^3_{\rm L} \times U(1)_{\rm B-L}^3$ is
\begin{equation}
Y^{\rm II}_{u,d} \sim \begin{bmatrix}
\epsilon_{\rm L}^2 & \epsilon_{\rm L}^2 \epsilon_q &  \epsilon_{\rm L}^2 \epsilon_q^2 \\
\epsilon_{\rm L} \epsilon_q & \epsilon_{\rm L} & \epsilon_{\rm L} \epsilon_q \\
\epsilon_q^2 & \epsilon_q & 1
\end{bmatrix},~
Y^{\rm II}_{e,\nu} \sim \begin{bmatrix}
\epsilon_{\rm L}^2 & \epsilon_{\rm L}^2 \epsilon_\ell &  \epsilon_{\rm L}^2 \epsilon_\ell^2\\
\epsilon_{\rm L} \epsilon_\ell & \epsilon_{\rm L} & \epsilon_{\rm L} \epsilon_\ell \\
\epsilon_\ell^2 & \epsilon_\ell & 1
\end{bmatrix}~,
\end{equation}
without affecting the Majorana mass matrix. This scenario satisfies the neutrino criteria in Eq.~\eqref{eq:condition} if
\begin{equation}
    \epsilon_{\rm L} = \epsilon_\ell~,
\end{equation}
and successfully generates hierarchical masses for quarks and charged leptons with singular values $(\epsilon_{\rm L}^2, \epsilon_{\rm L}, 1)$. In addition, it predicts a perturbative $V_{\rm CKM}$ matrix which is close to the unit matrix with $V_{us}\sim V_{cb}\sim  \epsilon_{\rm L} \epsilon_q$ and  $V_{ub} \sim V_{us} V_{cb}$. The predicted off-diagonal elements are arguably too small. However, in the minimal realization, where $\epsilon_q = \epsilon_{\rm L}^{1/3}$, this could be acceptable.

\subsection{Model C: $\SU(3)_{\rm C} \times \SU(2)^3_{\rm L} \times \U(1)^3_{\rm R} \times \U(1)^3_{\rm B-L}$}
\label{sec:modelIII}

While $\U(1)^3_{\rm R} \times \U(1)^3_{\rm B-L}$ yields a successful description of quark and charged-lepton Yukawa couplings, it does not satisfy the neutrino conditions in Eqs.~\eqref{eq:majorana} and \eqref{eq:condition}. However, this can be achieved via the deconstruction of the full electroweak group: $\SU(2)^3_{\rm L} \times \U(1)^3_{\rm R} \times \U(1)^3_{\rm B-L}$. This deconstruction implies
\begin{equation}
Y^{\rm III}_{u,d} \sim \begin{bmatrix}
\epsilon_{\rm L}^2 \epsilon_{\rm R}^2 & \epsilon_{\rm L}^2 \epsilon_{\rm R} \epsilon_q &  \epsilon_{\rm L}^2 \epsilon_q^2 \\
\epsilon_{\rm L} \epsilon_{\rm R}^2 \epsilon_q & \epsilon_{\rm L} \epsilon_{\rm R} & \epsilon_{\rm L} \epsilon_q \\
\epsilon_{\rm R}^2 \epsilon_q^2 & \epsilon_{\rm R} \epsilon_q & 1
\end{bmatrix},
\end{equation}
and
\begin{equation}
Y^{\rm III}_{e,\nu} \sim \begin{bmatrix}
\epsilon_{\rm L}^2 \epsilon_{\rm R}^2 & \epsilon_{\rm L}^2 \epsilon_{\rm R} \epsilon_\ell &  \epsilon_{\rm L}^2 \epsilon_\ell^2\\
\epsilon_{\rm L} \epsilon_{\rm R}^2 \epsilon_\ell & \epsilon_{\rm L} \epsilon_{\rm R} & \epsilon_{\rm L} \epsilon_\ell \\
\epsilon_{\rm R}^2 \epsilon_\ell^2  & \epsilon_{\rm R} \epsilon_\ell & 1
\end{bmatrix}~,
\end{equation}
while the Majorana mass matrix has the form in Eq.~\eqref{eq:majorana} with the replacement $\epsilon \to \epsilon_\ell \epsilon_R$. When
\begin{equation}
    \epsilon_{\rm L} = \epsilon_{\rm R} = \epsilon_\ell~,
\end{equation}
this texture produces neutrino anarchy while predicting quark and charged-lepton mass hierarchy,
with singular values $(\epsilon_{\rm L}^4, \epsilon_{\rm L}^2, 1)$  and the $V_{\rm CKM}$ matrix close to the unit matrix with $V_{us}\sim V_{cb}\sim  \epsilon_{\rm L} \epsilon_q$ and  $V_{ub} \sim V_{us} V_{cb}$. This relation works better when relating quark masses to the $V_{\rm CKM}$ elements than in Section~\ref{sec:modelII}.

At this point, it is tempting to ask what happens if, in addition, we also deconstruct $\SU(3)_{\rm C}$ such that the full underlying gauge group is
\begin{equation}
    {\rm PS}^{(1)}\times {\rm PS}^{(2)} \times {\rm PS}^{(3)}~,
\end{equation}
as initially conjectured in~\cite{Bordone:2017bld}. Staying at the level of $\SU(3)^3_{\rm C} \times \SU(2)^3_{\rm L} \times \U(1)^3_{\rm R} \times \U(1)^3_{\rm B-L}$, we simply assume $\phiH^q$ in Table~\ref{tab:scalars2} to be a $({\bf 3}, {\bf \bar 3})$ under $\SU(3)^{(1)}_{\rm C} \times \SU(3)^{(2)}_{\rm C}$ and the same for $\phiL^q$ under $\SU(3)^{(2)}_{\rm C} \times \SU(3)^{(3)}_{\rm C}$. Here, however, we need to have $\phi^\ell_{ij}$ fields as well. In fact, $\phi^q_{ij}$ and $\phi^\ell_{ij}$ unify into a pair of bifundamentals $\Sigma_{ij}$ of $\SU(4)^3 \supset \SU(3)^3_{\rm C} \times \U(1)^3_{\rm B-L}$. Remarkably, this model predicts exactly the flavor structure of Model C!

\section{Deconstructing the inverse seesaw} 
\label{sec:typeII}

The inverse seesaw~\cite{Wyler:1982dd, Mohapatra:1986bd} belongs to a class of symmetry-protected seesaw mechanisms that predict lighter sterile neutrinos with larger couplings. This provides an interesting target for probing the origin of neutrino masses in ongoing and upcoming experiments. In addition to left-chiral $\ell_i$ and right-chiral $\nu_i$, the model is extended by (left-chiral) gauge singlets $S_i$, where for concreteness, we consider three flavors ($i=1,2,3$). Let us define $n_i = (\ell^1_i, \nu^c_i, S_i)$. Here, $\ell^1$ stands for the neutral component of the lepton doublet and $c$ is for the charge conjugation. The neutral lepton mass matrix ($9\times9$) is
\begin{equation}\label{eq:iss}
-\mathcal{L} \supset \frac{1}{2} \bar n_i M^{ij}_{\nu} n_j^{c}, \quad
M_\nu = \begin{bmatrix}
0 & M_D & 0 \\
M_D^T & 0 & M_R^T \\
0 & M_R & \mu_S
\end{bmatrix}~,
\end{equation}
The global lepton number is an approximate symmetry softly broken by $\mu^{i j}_S$. For $\mu_S \ll M_D < M_R$,\footnote{The PMNS non-unitarity sets limit on $M_D / M_R$, see e.g.~\cite{Antusch:2016ejd}.} the active neutrino mass matrix takes the form
\begin{equation} \label{eq:issmnu}
    m_\nu \approx M_D M_R^{-1} \mu_S (M_R^{-1})^T M^T_D~.
\end{equation}
For the same arguments presented in Section~\ref{sec:typeI}, a deconstruction of the SM gauge symmetry cannot reconcile hierarchical charged leptons with anarchic neutrinos. Therefore, we are again led to consider the $\SU(3)_{\rm C} \times \SU(2)_{\rm L} \times \U(1)_{\rm R} \times \U(1)_{\rm B-L}$ gauge symmetry. 
Introducing the scalar field $\chi \sim ({\bf 1}, {\bf 1}, -1/2, +1)$ that breaks the two $\U(1)$'s down to the SM hypercharge, allow us to write
\begin{equation}
    -\mathcal{L} \supset Y^{ij}_R \bar S_i \chi \nu_j~, \quad M^{ij}_R = Y^{ij}_R \langle \chi \rangle ~.
\end{equation}
Note that the Majorana mass term for $\nu_i$ is absent at the renormalizable level. As before, $M^{ij}_D = Y^{ij}_\nu \langle H \rangle$ stems from $ \mathcal{L} \supset - Y^{ij}_\nu \bar \ell_i \tilde H \nu_j$. The remaining renormalizable operator $\mathcal{L} \supset - \frac{1}{2} \mu_S^{ij} \bar S^c_i S_j$ finally justifies Eq.~\eqref{eq:iss}. 

Since the $S_i$ are gauge singlets, the natural expectation for $\mu_S^{ij}$ is flavor anarchy. On the other hand, flavor deconstruction imprints hierarchical patterns in $ Y_\nu $ and $Y_{R}$. When put together, Eq.~\eqref{eq:issmnu} requires $Y_\nu Y^{-1}_R$ to be an anarchic matrix for $m_\nu$ also to be anarchic. Under $\U(1)_{\rm R}$ or $\U(1)_{\rm B-L}$ deconstruction, $Y_R$ takes the form of Eq.~\eqref{eq:U1Rdec}. As a result, $Y_\nu$ should satisfy the same condition as in Eq.~\eqref{eq:condition}, see Appendix~\ref{app:ISS}. 

Since the condition in Eq.~\eqref{eq:condition} is unchanged, the same gauge symmetries that worked for the high-scale seesaw discussed in Sections~\ref{sec:modelI} and \ref{sec:modelII} are also applicable in the inverse seesaw case. This is not surprising given the additional fermions we have introduced (the $S_i$), being gauge singlets do not participate in the flavor deconstruction. The observed neutrino mass splittings suggest $\mu \sim 0.1\,{\rm eV} \langle \chi \rangle^2 / \langle H \rangle^2$. Therefore, the major advantage of the inverse seesaw is that the deconstruction can easily take place at the TeV scale, which is consistent with finite naturalness (see also~\cite{FileviezPerez:2023rxn}). Detailed phenomenological investigations are left for future studies.


\section{Conclusions} 
\label{sec:conc}

Seemingly anarchic neutrino flavor structure, when contrasted with the hierarchies observed in the charged fermion sector, poses a significant challenge for many flavor models, especially those involving flavor deconstruction. The SM gauge group's deconstruction fails to reconcile both phenomena simultaneously. However, as demonstrated in this letter, minimal extensions where right-handed neutrinos are not gauge singlets allow for reconciliation, enabling the correct generation of charged fermion mass and mixing hierarchies while predicting anarchy for the light-active neutrinos. Focusing on the type-I and inverse seesaw mechanisms, we have conducted a comprehensive survey of possible options and identified promising models for further study. While these models strictly lead to neutrino anarchy, other deconstruction models with moderate tuning should not be disregarded.

\section*{Acknowledgments}

We thank Joe Davighi for his useful comments on the manuscript. This work has received funding from the Swiss National Science Foundation (SNF) through the Eccellenza Professorial Fellowship ``Flavor Physics at the High Energy Frontier,'' project number 186866.  In addition, this project has received funding from the European Research Council (ERC) under the European Union’s Horizon 2020 research and innovation program under grant agreement 833280 (FLAY) and by the SNF under contract 200020\_204428.

\appendix 
\renewcommand{\thesection}{\Alph{section}}
\renewcommand{\thesubsection}{\Alph{section}.\arabic{subsection}}
\setcounter{section}{0}

\section{Type-I seesaw: Hierarchy cancellation}
\label{app:typeI}

Consider the matrices
\begin{equation}
X_{\rm M} = \begin{pmatrix}
x_{11} \epsilon^2_1 \epsilon_2^2 & x_{12} \epsilon_1 \epsilon_2^2 & x_{13} \epsilon_1 \epsilon_2 \\
x_{12} \epsilon_1 \epsilon_2^2 & x_{22} \epsilon_2^2 & x_{23} \epsilon_2 \\
x_{13} \epsilon_1 \epsilon_2 & x_{23} \epsilon_2 & x_{33}
\end{pmatrix}\,, \quad 
Y_{\rm D} = \begin{pmatrix}
\epsilon_1 \epsilon_2 & 0 & 0 \\
0 & \epsilon_2 & 0 \\
0 &  0 & 1
\end{pmatrix}\,.
\end{equation}
The following relation holds 
(with no approximations):
\begin{equation}
    Y_{\rm D} X^{-1}_{\rm M} Y_{\rm D}^T = \begin{pmatrix}
x_{11} & x_{12} & x_{13} \\
x_{12} & x_{22}  & x_{23}  \\
x_{13} & x_{23} & x_{33}
\end{pmatrix}^{-1}~.
\end{equation} 
For $x_{ij} \sim \mathcal{O}(1)$ and $\epsilon_i \ll 1$, this provides a mechanism to generate an anarchic mass matrix for active neutrinos starting from hierarchical Yukawa couplings in the type-I seesaw mechanism.

\section{Inverse seesaw: Hierarchy cancellation} 
\label{app:ISS}

Consider the matrices
\begin{equation}
Y_{\rm D} = \begin{pmatrix}
y_{11} \epsilon_1 \epsilon_2  & y_{12} \epsilon_2 & y_{13} \\
y_{21} \epsilon_1 \epsilon_2 & y_{22} \epsilon_2 & y_{23} \\
y_{31} \epsilon_1 \epsilon_2  & y_{32} \epsilon_2 & y_{33}
\end{pmatrix}\,, \quad 
Y_{R} = \begin{pmatrix}
\epsilon_1 \epsilon_2 & 0 & 0 \\
0 & \epsilon_2 & 0 \\
0 &  0 & 1
\end{pmatrix}\,.
\end{equation}
The following relation holds 
(with no approximations):
\begin{equation}
    Y_{\rm D} Y^{-1}_{\rm R} = \begin{pmatrix}
y_{11}  & y_{12}  & y_{13} \\
y_{21}  & y_{22} & y_{23} \\
y_{31}  & y_{32} & y_{33}
\end{pmatrix}~.
\end{equation} 
For $y_{ij} \sim \mathcal{O}(1)$ and $\epsilon_i \ll 1$, this provides a mechanism to generate an anarchic mass matrix for active neutrinos in the inverse seesaw mechanism.

\bibliographystyle{JHEP}
\bibliography{refs.bib}

\end{document}